\documentclass[12pt, draftcls, onecolumn]{IEEEtran}

\usepackage[cmex10]{amsmath} 
\usepackage{amsfonts,amssymb,amsthm}
\usepackage{graphicx}
\usepackage{url}
\usepackage{balance}
\usepackage{cite}

\usepackage{enumerate}
\usepackage{bbm}
\usepackage{color}
\usepackage{soul}

\makeatletter
\def\blfootnote{\xdef\@thefnmark{}\@footnotetext}
\makeatother

\newcounter{mytempeqcounter}

\allowdisplaybreaks
\allowbreak

\makeatletter%
\if@twocolumn%
\newcommand{\Figwidth}{\columnwidth}%

\else
\newcommand{\Figwidth}{4in}%

\fi%
\makeatother%

\IEEEoverridecommandlockouts

\begin{document}

\title{Compress-and-Forward via  Multilevel Coding and Trellis Coded Quantization}
\author{Heping Wan, Anders H{\o}st-Madsen, {\em Fellow, IEEE}, and Aria Nosratinia, {\em Fellow, IEEE}
\thanks{This work of H. Wan and A. Nosratinia was supported by the grant 1711689 from the National Science Foundation. The work of A. Host-Madsen was supported by the grant 1923751 from the National Science Foundation.}
\thanks{H. Wan and A. Nosratinia are with the Department of Electrical Engineering, The University of Texas at Dallas, Richardson, TX, USA,Email: Heping.Wan@utdallas.edu, aria@utdallas.edu A. Host-Madsen is with the Department of Electrical Engineering, University of Hawaii, Manoa, Honolulu, HI, USA Email: ahm@hawaii.edu}
}

\maketitle
\date{}

\begin{abstract}
Compress-forward (CF) relays can improve communication rates even when the relay cannot decode the source signal. Efficient implementation of CF is a topic of contemporary interest, in part because of its potential impact on wireless technologies such as cloud-RAN. There exists a gap between the performance of CF implementations in the high spectral efficiency regime and the corresponding information theoretic achievable rates. We begin by re-framing a dilemma causing this gap, and propose an approach for its mitigation. We utilize trellis coded quantization (TCQ) at the relay together with multi-level coding at the source and relay, in a manner that facilitates the calculation of bit LLRs at the destination for joint decoding. The contributions of this work include designing TCQ for end-to-end relay performance, since a distortion-minimizing TCQ is suboptimum. The reported improvements include a 1dB gain over prior results for PSK modulation.
\end{abstract}

\section{Introduction}
\label{sec:intro}


This paper addresses compress-forward (CF) relaying under high-spectral efficiency (coded modulation), where implementations of CF show a performance gap to the best known (information theoretic) achievable rates.  The nature of the difficulties is succinctly explained as follows. The best theoretical achievable rates employ either Wyner-Ziv compression at the relay, or joint decoding at the destination~\cite{Kramer_short, Zhong_withoutWynerZiv}. Since practical Wyner-Ziv coding implementations have had some difficulty in approaching the theoretical limits, joint decoding at the destination has been pursued~\cite{Nagpal_QMF_Allerton, Nagpal_QMF_Selected} via iterative decoding. Iterative decoding requires bit-level LLRs, which are easy to obtain under scalar quantization but difficult with Vector Quantization (VQ). Thus, joint decoding approaches so far have utilized scalar quantization and given up on the shaping gain that is only provided by vector quantization. The current paper addresses this issue and, via a new approach, improves the performance of CF implementations. 

We begin with a brief survey of literature.
%
In~\cite{A_layered_cf} the relay quantizes soft information of received values followed by a multi-level coding with convolutional codes for each level, which serves as both Slepian-Wolf compression as well as error protection. 
The works~\cite{Uppal_bpsk,Uppal_MLC_Globecom,Blasco_polar_Ic}, feature half-duplex relay, Wyner-Ziv coding, and successive decoding at destination. \cite{Uppal_bpsk,Uppal_MLC_Globecom} employ LDPC codes at the source and irregular-repeat accumulation (IRA) codes at the relay. \cite{Blasco_polar_Ic} uses a nested construction of polar codes at the relay. Liu et al.~\cite{cf_superposition} implements Wyner-Ziv coding via a superposition structure.

An alternative strategy maps the quantized sequence to a relay codeword {\em without} Wyner-Ziv encoding~\cite{Infor_flow,Kramer_short,Zhong_withoutWynerZiv}, but with joint decoding at destination. This strategy has the same achievable rate as Wyner-Ziv encoding with successive cancellation decoding. The following works utilized scalar quantization and joint decoding.
%
%
Chakrabarti et al.~\cite{PracticalQuantizerDesign} designed quantizers by maximizing the mutual information between the source and the quantizer output instead of minimizing the distortion. Nagpal {\em et al.}~\cite{Nagpal_QMF_Allerton} used LDPC codes, and Nagpal {\em et al.}~\cite{Nagpal_QMF_Selected} used Low Density Generator Matrix (LDGM) codes to map the scalar quantized symbols to codewords. These binary modulations were extended to coded modulation via bit-interleaved coded modulation (BICM). The present authors employed this approach in~\cite{CF_Multilevel} with scalar quantization, multilevel coding, and LDPC codes, and joint decoding.

As mentioned earlier, scalar quantization is easier to manage in soft decoding, but sacrifices the shaping gain. A general vector quantizer (VQ) captures the shaping gain in principle, but makes the calculation of bit LLRs intractable. To resolve this tension, we utilize trellis coded quantization (TCQ)~\cite{TCQ}. 
In this technique, trellis coding captures shaping gain while likelihood exchange is enabled by the BCJR algorithm~\cite{BCJR_TrellisSourceCoding}.

We utilize this technique for the design of coded modulation for full-duplex compress-forward relays in the AWGN channel. We employ multilevel coding (MLC)~\cite{Imai_MLC_It} as a convenient and flexible method for the implementation of coded modulation in this multi-node network. We analyze the proper assignment of rates to the component level-wise binary codes for near-optimal error protection; this is in contrast with~\cite{Nagpal_QMF_Allerton,Nagpal_QMF_Selected} which were limited to the same rate at all levels. Part of the contribution of this work is a practical design of the TCQ that targets end-to-end relay performance, since a distortion-minimizing TCQ proved insufficient. Quantizer optimization, and the related computation of input-output distributions for the soft decoding of TCQ, are elucidated.
We use LDPC codes for error protection at the source, and map the quantized sequence at the relay to a codeword whose parity bits are transmitted. A joint iterative decoding graphical model and corresponding information exchange algorithm are described for the joint decoding at the destination. 

In summary, contributions of this letter include: TCQ for the CF relay to achieve shaping gain while allowing LLR exchange at the iterative decoder,
MLC per-level rate analysis for CF relaying, iterative joint decoding, and demonstrating the gains of the proposed approach.


\section{Setup, Notation, and Preliminaries\protect\footnote{This section appears here for completeness and follows~\protect\cite{CF_Multilevel} in its entirety.}}
\label{sec:preli}

In the point-to-point channel, MLC is implemented by splitting the data stream into $m$ bit-streams for a $2^m$-ary constellation. Each sub-stream $i\in\{1,2,..., m\}$ is encoded independently. At each time instance, the outputs of the (binary) encoders are combined to construct vector $[A_1, A_2,...A_m]$ which is then mapped to a constellation point $X$ and transmitted. At the destination, $Y$ is observed. The channel is described by conditional distribution $p(y|x)$. The mutual information between the input and output is given by 
\begin{align}
I(X;Y)=I(A_1, A_2,...,A_m;Y)=\sum\limits_{i=1}^{m}I(A_i;Y|A^{i - 1}), \label{eq:p2pmlc}
\end{align}
where $A^{i - 1} \triangleq [A_1, A_2,...,A_{i - 1}]$ with $A_0$ representing a constant, and the chain rule for the mutual information and a one-to-one relationship between $X$ and $[A_1, A_2,...,A_m]$ are used. Equation \eqref{eq:p2pmlc} suggests a multistage decoding, and the original channel is decomposed into $m$ levels where the information bits of level $i$ is recovered using channel observations and the output of decoders of preceding levels. Therefore, the rate assigned to level $i$ should be less than or equal to $I(A_i;Y|A^{i - 1})$. 

\begin{figure}[h]
\centering
\includegraphics[width=2.5in]{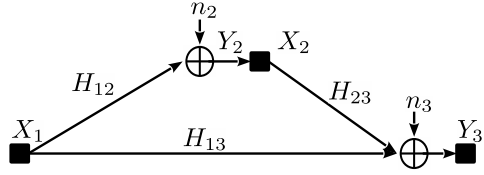}
\caption{Full-duplex Gaussian relay channel}
\label{fig:channel model}
\end{figure}
The three-node AWGN full-duplex relay channel is shown in Fig.~\ref{fig:channel model}. The source transmits $X_1\in{\bf \mathcal{A}}$ satisfying power constraint $P_s$ to the destination and relay. The relay transmits $X_2\in{\bf \mathcal{B}}$ satisfying power constraint $P_r$ to the destination. ${\bf \mathcal{A}}$ and ${\bf \mathcal{B}}$ denote the constellation alphabets of $X_1$ and $X_2$. ${n_2}$ and ${n_3}$ are AWGN with zero mean and variance $N_2$ and $N_3$. The received signals at the relay and destination are
\begin{align}
{Y_2} & = {H_{12}}{X_1} + {n_2},\label{eq:receivedy2}\\
{Y_3} & = {H_{13}}{X_1} + {H_{23}}{X_2} + {n_3},\label{eq:receivedy3}
\end{align}
 where $H_{13}$, $H_{12}$ and $H_{23}$ are channel coefficients as illustrated in Fig.~\ref{fig:channel model}. 

In block $t$, the source maps one of $2^{nR}$ messages to a length-$n$ codeword ${\mathbf X}^{(t)}_1$, and emits ${\mathbf X}^{(t)}_1$ to the destination and relay. Due to causality, the relay quantizes the sequence received in block $t - 1$. $\Tilde{\mathbf {Y}}^{(t-1)}_2$ of length $n$ denotes the quantized binning of ${\mathbf Y}^{(t-1)}_2$. Then $\Tilde{\mathbf {Y}}^{(t-1)}_2$ is mapped to a length-$n$ codeword ${\mathbf X}^{(t-1)}_2$ and transmitted to the destination. The destination uses the joint decoding to recover the source message sent at block $t-1$. The achievable rate is~\cite[Section 16.7]{ElGamalKim_NIT}
\begin{align}
R <\max& \, I(X_1;\Tilde{Y}_2,Y_3|X_2),\label{eq:cf_achi_rate}
\end{align}
subject to $\quad I(Y_2;\Tilde{Y}_2|X_2,Y_3) \leq I(X_2;Y_3),$ where the maximum is over joint distributions \newline $p(x_1)p(x_2)p(\Tilde{y}_2|x_2,y_2)$.

When MLC is implemented in the CF relaying under $2^m$-QAM/PSK modulation for both source and relay, the data stream at the source and the quantized binning at the relay are split into $m$ bit-streams. Each of $m$ binary bit-streams at the source/relay is encoded independently. Due to one-to-one relationships between $X_1\leftrightarrow [A_1, A_2,...,A_m]$, $\Tilde{Y}_2 \leftrightarrow [B_1, B_2,...,B_m]$, and $X_2 \leftrightarrow [C_1, C_2,...,C_m]$, multistage decoding will induce a decomposition of the original channel, and the achievable rate in \eqref{eq:cf_achi_rate} can be expressed as
\begin{align}
R\leq \max \, \sum\limits^{m}_{i=1}I(A_i; B^m,Y_3|C^m,A^{i-1}).\label{eq:level_achi_rate}
\end{align}
subject to the constraint:
\begin{equation}
\sum\limits^{m}_{i=1}I(Y_2;B_i|C^m,Y_3,B^{i-1}) \leq \sum\limits^{m}_{i=1}I(C_i;Y_3|C^{i-1}) \; .
\end{equation}
The rate assigned to each level is denoted $R_i$, requiring
\begin{equation}
R_i \leq I(A_i; B^m,Y_3|C^m,A^{i-1})
\label{eq:levelwiseRate}
\end{equation}

\section{Multilevel Realization of CF relaying}
\label{sec:multilevelCF}
The message to be transmitted is partitioned into $m$ bit-streams. Each bit-stream is independently encoded with binary LDPC code of rate $R_i, i=1,\dots,m$, where $R_i$ should satisfy \eqref{eq:levelwiseRate}. At each time instance, binary vector $[A_1, A_2,...A_m]$ is mapped to a symbol of the transmitted sequence ${\mathbf X}_1$ by a $2^m$-QAM/PSK modulator. At the relay, the quantized binning $\Tilde{\mathbf {Y}}_2$ is split into $m$ bit-streams. Each bit-stream is encoded through a systematic rate-$1/2$ binary LDPC code. The parity bits of the codeword are combined via a $2^m$-QAM/PSK modulator, to construct ${\bf X}_2$ which is transmitted to the destination at the next block. 

\subsection{Quantization at the Relay}
\label{sec:qua}

The key impediment to CF performance has been the implementation of compression at the relay. Wyner-Ziv implementations have not approached fundamental limits, and joint decoding has concentrated on scalar quantization to facilitate LLR calculation, thus losing shaping gain of the quantizer at the relay. We provide a new solution via TCQ that reconciles this issue by providing shaping gain as well as feasible LLR calculation for joint decoding.


Trellis source coding aims to compress a sequence $\textbf{y} = [y_1,y_2,\dots,y_n]$ sampled from a memoryless source with marginal distribution $p(y)$ to $\Tilde{\mathbf{y}} = [\Tilde{y}_1,\Tilde{y}_2,\dots,\Tilde{y}_n]$ at a rate of $\Tilde{R}$. The reconstruction values are $\hat{y} \in {\bf \cal \hat{Y}}$. Trellis source coding is to find a path determined by $\Tilde{\mathbf{y}}$ which generates a reconstruction sequence $\hat{\mathbf{y}}$ with minimum distortion.

TCQ~\cite{TCQ} is a practical implementation of trellis source coding that was inspired by trellis coded modulation (TCM)~\cite{TCM_Ungerboeck}. In TCM, to send $m$ information bits in each signaling interval, the traditional $2^m$- point signal constellation is expanded to $2^{m+1}$ points and partitioned into $2^{m}$ subsets according to Ungerboeck's set partitioning idea. $m$ input bits are passed through a rate $m/(m+1)$ trellis encoder among which $m-1$ bits are used to select the set partition, and the remaining bit determines the symbol from the set partition. 

\subsection{Obtaining $p(\hat{y}_2|y_2)$ and Quantizer Optimization}
\label{Sec:QuantizerOptimization}
For iterative decoding at the destination, one needs a soft version of TCQ, for which a logical candidate is the Bahl-Cocke-Jelinek-Raviv (BCJR) algorithm. To implement the BCJR, the (marginal) conditional distribution $p(\hat{y}_2|y_2)$ is needed \cite{BCJR_TrellisSourceCoding}, where $\hat{y}_2 \in {\bf\cal \hat{Y}}_2$ represents reconstruction value and $|{\bf\cal \hat{Y}}_2|=2^{m+1}$. Unfortunately, this distribution is not readily available and its calculation is not straight forward. In this subsection, we highlight the different approaches for calculating this value, and present the technique that we utilized, which to our experience produces the best results.



Rate-distortion theory~\cite{Shannon_RateDistortion} suggests an optimal source encoder coming from the solution of the following minimization
\[
R(D)=\underset{p(\hat{y}|y):E\{d(y,\hat{y})\}\leq D}{\min}I(Y,\hat{Y}),
\]
where the minimization is over all conditional distributions $p(\hat{y}|y)$ for which the joint distribution $p(\hat{y},y)=p(y)p(\hat{y}|y)$ satisfies the expected distortion constraint $D$, and $d(y,\hat{y})$ is the distortion measure function. $p(\hat{y}|y)$ can be interpreted as the probability that a given source value $y$ is represented by a reconstruction value $\hat{y}$.

With the resulting $p(\hat{y}|y)$ obtained from the above optimization, the encoding procedure is equivalent to finding a $\hat{y}(n)$ that maximizes $p(\hat{y}(n),\mathbf{y})$. For trellis source coding, the BCJR algorithm~\cite{BCJR_TrellisSourceCoding} can be used to find the reconstruction sequence $\hat{\mathbf {y}}$  and the corresponding compressed bitstream $\Tilde{\mathbf {y}}$.

However, minimizing distortion is not necessarily a good strategy for CF relaying, because the ultimate goal is for relaying to give maximal help in decoding at destination. Counterintuitively, this is not always the same as low distortion, which only preserves the fidelity of a reconstruction of ${\mathbf y}_2$. Our simulations indicate that Lloyd-Max reconstruction values and $p(\hat{y}_2|y_2)$ obtained from rate-distortion optimization together produce mediocre performance.\footnote{This issue has also been pointed out by others, e.g., Chakrabarti et al.~\cite{PracticalQuantizerDesign} produced a comparison in which a quantizer with worse mean-squared distortion resulted in better relay performance.} This seems to suggest jointly designing $p(\hat{y}_2|y_2)$ and the trellis structure, but this is too unwieldy for practical quantizer design. 
Our results are obtained by choosing a good trellis from the literature and determining quantizer boundaries by maximizing the CF (end-to-end) achievable rate expression, which is related to the quantization boundaries through the conditional distribution $p(\hat{y}_2|y_2)$. The boundaries were determined via a simple discrete search that maintained the symmetry of the quantizer and utilized the trellis and the reconstruction values in Fig.~\ref{fig:Reconstruction values}.

\subsection{Decoding}
\label{sec:de}

As mentioned earlier, the destination decodes the source signal as well as the quantized signal at the relay; information theory arguments show that this joint decoding permits the relay to avoid dirty-paper type encoding without a rate penalty. The joint decoder is implemented iteratively via a graphical model for the exchange of information within, and between, the component LDPC tanner graphs.


The desination begins by seeking a codeword ${\mathbf x}_1$ according to maximum {\em a posteriori} probability $p{({\mathbf x}_1|{\mathbf y}_3})$; this search is performed via a bit-wise MAP decoder with decoding rule:
\[
\hat{x}_{1,j}=\underset{x_{1,j}\in{\mathcal{A}}}{\arg\max}\sum\limits_{\sim x_{1,j}}p({\mathbf x}_1|{\mathbf y}_3),
\]
for all $j = 1,...,n,$. Recall that $\hat{\mathbf y}_2$ are the quantized values while $\Tilde{\mathbf y}_2$ is the compressed bit sequence. These two representations are equivalent given the quantization strategy, however, $\Tilde{\mathbf y}_2$ is a more  convenient representation for the relay's channel encoder as well as for iterative decoding at destination. $p({\mathbf x}_1|{\mathbf y}_3)$ can be decomposed as follows:
\begin{align}
& p({\mathbf x}_1|{\mathbf y}_3) = \frac{1}{p({\mathbf y}_3)} \sum\limits_{\Tilde{\mathbf{y}}_2,{\mathbf x}_2}p({\mathbf x}_1,{\mathbf x}_2,\Tilde{\mathbf {y}}_2,{\mathbf y}_3),\label{eq:map6}
\end{align}
where \begin{align}
&p({\mathbf x}_1,{\mathbf x}_2,\Tilde{\mathbf {y}}_2,{\mathbf y}_3) = p({\mathbf y}_3|{\mathbf x}_1,{\mathbf x}_2,\Tilde{\mathbf{y}}_2)p({\mathbf x}_1,{\mathbf x}_2,\Tilde{\mathbf{y}}_2),\label{eq:map7}\\
& \overset{{(a)}}= p({\mathbf y}_3|{\mathbf x}_1,{\mathbf x}_2)p({\mathbf x}_1,{\mathbf x}_2,\Tilde{\mathbf{y}}_2),\label{eq:map8}\\
& =p({\mathbf y}_3|{\mathbf x}_1,{\mathbf x}_2)p({\mathbf x}_2|{\mathbf x}_1,\Tilde{\mathbf{y}}_2)p(\Tilde{\mathbf{y}}_2|{\mathbf x}_1)p({\mathbf x}_1),\label{eq:map9}\\
& \overset{{(b)}}=p({\mathbf y}_3|{\mathbf x}_1,{\mathbf x}_2)p({\mathbf x}_2|\Tilde{\mathbf{y}}_2)p(\Tilde{\mathbf{y}}_2|{\mathbf x}_1)p({\mathbf x}_1),\label{eq:map10}\\
& \overset{{(c)}} \propto p({\mathbf y}_3|{\mathbf x}_1,{\mathbf x}_2)
\mathbbm{1}\{{\mathbf x}_2=\mathcal{C}_R(\Tilde{\mathbf{y}}_2)\}p(\Tilde{\mathbf{y}}_2|{\mathbf x}_1)\mathbbm{1}\{{\mathbf x}_1\in{\mathbf \mathcal{C}_S}\},
\label{eq:map11}
\end{align}
where $(a)$ follows because ${\mathbf y}_3$ is only dependent on ${\mathbf x}_1,{\mathbf x}_2$, $(b)$~is due to Markov ${\mathbf x}_1 \leftrightarrow \Tilde{\mathbf{y}}_2 \leftrightarrow {\mathbf x}_2$, and $(c)$~is due to $p({\mathbf x}_1)=\frac{\mathbbm{1}\{{\mathbf x}_1\in{\mathbf \mathcal{C}_S}\}}{2^{nR}}$ being a uniform distribution over the elements of the source codebook $\mathcal{C}_S$, and $p({\mathbf x}_2|\Tilde{\mathbf{y}}_2)=\mathbbm{1}\{{\mathbf x}_2=\mathcal{C}_R(\Tilde{\mathbf{y}}_2)\}$, i.e., the output of error-control coding at the relay $\mathcal{C}_R$ being a deterministic function of its input.

\begin{figure}[h]
\centering
\includegraphics[width=2.in]{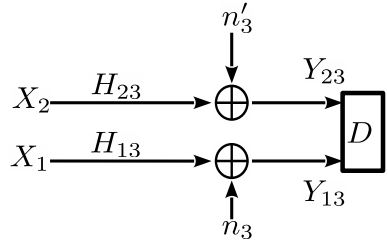}
\caption{The simplified destination part of the channel model}
\label{fig:simplified channel model}
\end{figure}
During block $t$, the destination receives a version of ${\mathbf X}^{(t-1)}_2$ that is contaminated with noise and interference, which carries information about ${\mathbf X}^{(t-1)}_1$, which is independent of ${\mathbf X}^{(t)}_1$. This independence allows successive interference cancellation, simplifying the channel model at destination~\cite{Nagpal_QMF_Allerton}:
\begin{align}
& {Y}_{23} = {H_{23}}{X_2} + {n'_3},\label{eq:y23}\\
& {Y}_{13} = {H_{13}}{X_1} + {n_3},\label{eq:y13}
\end{align}
which is shown in Fig.~\ref{fig:simplified channel model}. At block t, the destination jointly decodes ${\mathbf X}^{(t-1)}_1$ and ${\mathbf X}^{(t-1)}_2$ on the basis of ${\mathbf Y}^{(t)}_{23}$ and ${\mathbf Y}^{(t-1)}_{13}$ where the processing of ${\mathbf Y}^{(t)}_{23}$ treats ${\mathbf X}^{(t)}_1$ as zero-mean Gaussian noise with power $H^2_{12}P_s$, and ${\mathbf Y}^{(t-1)}_{13}$ is processed by subtracting ${\mathbf X}^{(t-2)}_2$ decoded during block $t-1$ from ${\mathbf Y}^{(t-1)}_3$. This operation can decompose $Y_3$ to two links $Y_{13}$ and $Y_{23}$ with independent additive white Gaussian noise $n'_3$ of variance $N_3+H^2_{12}P_s$ and $n_3$ of variance $N_3$. Using $p(y_3|x_1.x_2)=p(y_{31}|x_1)p(y_{32}|x_2)$, for LLR calculation at destination, \eqref{eq:map11} can be rewritten as
\begin{align}
\prod\limits_{j=1}^{n} & p(y_{13,j}|x_{1,j}) \prod\limits_{k=1}^{n}p(y_{23,k}|x_{2,k})\nonumber\\
& \cdot \mathbbm{1}\{{\mathbf x}_1\in{\mathbf \mathcal{C}_S}\}p(\Tilde{\mathbf{y}}_2|{\mathbf x}_1)\mathbbm{1}\{{\mathbf x}_2=\mathcal{C}_R(\Tilde{\mathbf{y}}_2)\},\label{eq:map14}
\end{align}

\begin{figure}[t]
\centering
\includegraphics[width=4.in]{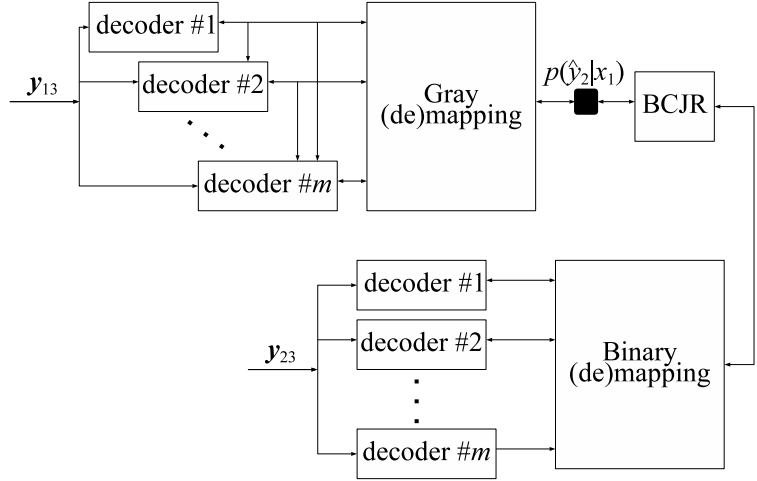}
\caption{The joint iterative decoding graphical model}
\label{fig:Tanner graph}
\end{figure}

Enabled by this decomposition, joint decoding can be performed  by a repeated application of two LDPC decoders whose cost function is based on $p(y_{13}|x_1)$ and $p(y_{23}|x_2)$, as well as enforcing the relationship between $x_1$ and $x_2$ through the term $p(\Tilde{\mathbf y}_2|\mathbf{x}_1)$ together with the relay LPDC coding ${\mathbf x}_2={\mathcal C}_R (\tilde{\mathbf y}_2)$. Using the destination observations $y_{13}$ and $y_{23}$ (after successive interference cancellation) a three-way iterative decoding is performed whose information exchange is described by Fig.~\ref{fig:Tanner graph}, where each LDPC decoder block represents a Tanner graph.


The two LDPC decoder components follow well-known principles, therefore their internal structure is omitted in Fig.~\ref{fig:Tanner graph} for simplicity.  The iterative decoding also needs propagating soft information through $p(\Tilde{\mathbf y}_2|\mathbf{x}_1)$, which is not straight forward, therefore we break it into two components.  Considering the Markov chain $\mathbf{x}_1 \leftrightarrow \hat{\mathbf{y}}_2 \leftrightarrow \tilde{\mathbf{y}}_2$, we have $p({\Tilde{\mathbf{y}}}_2|{\mathbf{x}}_1)=\sum\limits_{\hat{{\bf y}}_2} p({\Tilde{\mathbf{y}}}_2|\hat{\mathbf{y}}_2)p(\hat{\mathbf{y}}_2|{\mathbf{x}}_1)$. The information exchange across the trellis, shown by $p(\Tilde{\mathbf y}_2|\hat{\mathbf y}_2)$, is calculated using the BCJR algorithm. The information exchange between ${\mathbf x}_1$ and $\hat{\mathbf y}_2$ is through $p(\hat{y}_2|x_1)$.
%
%
This marginal distribution of quantized values is described as a weighted sum since TCQ employs several scalar quantizers, whose number we denote with $K$. The quantizer choice at each sample has a probability ${\mathbb P}(Q_i|x_1)$. The overall marginal distribution is given by:  
\begin{equation}
p(\hat{y}_2|x_1) =  \sum_{i=1}^{K}\; {\mathbb P}(Q_i|x_1) \!\!\!\!\int\limits_{\hat{y}_2=Q_i(y_2)}\! p(y_2|x_1)\;dy_2
\label{eq:scalarQ}
\end{equation}
The probability ${\mathbb P}(Q_i|x_1)$ at each sample is driven by the contamination of $x_1$ with noise prior to quantization, as well as the action of the trellis. The overall effect is difficult to model, therefore we obtain an empirical estimate for ${\mathbb P}(Q_i|x_1)$ by a monte carlo simulation of the channel $x_1\rightarrow y_2$, applying the TCQ, and collecting the empirical frequency of occurrence of each quantizer $Q_i$ conditioned on $x_1$. 

The destination begins by initializing $p(y_{13})$ and $p(y_{23})$ according to channel observations. Then, we update the soft estimate of $x_1$ at the destination, via executing the sum-product algorithm for ${\cal C}_S$. From this, an updated soft estimate for $\hat{y}_2$ is obtained via ~\eqref{eq:scalarQ}. Then, the BCJR algorithm, induced by the compression trellis, is used to update the soft estimate of $\tilde{y}_2$. Then, another sum-product is executed utilizing the observation $y_{23}$ and the structure of the relay-destination LDPC code, to update $\tilde{y}_2$. We then repeat backward, updating $\hat{y}_2$ and $x_1$, and then another sum-product iteration on $x_1$. 

To summarize, the process consists of two sum-product algorithms operating on the source LDPC code and the relay LDPC code, and information being exchanged between them via the method mentioned above. Our algorithm performs one bit-to-check and one check-to-bit operation in each sum-product before propagating the information to the other LDPC decoder. At the end of the process, we perform a multistage hard decoding of the MLC.


\section{Simulations}
\label{sec:sim}

This section presents simulation results supporting the findings of this letter. The following values underlie the simulations: $N_3=1$, $N_2=8$, $H_{13}=1$, $H_{12}=2$, and $H_{23}=11$. Furthermore, in all experiments the source and the relay transmitters emit the same power. The reported SNRs for simulations are generated by fixing the noise power and varying the power of the source/relay.


Error control coding is achieved via DVB-S2 LDPC codes with block length 64,800. At the relay, the first 32,400 quantized bits are encoded by a rate-1/2 DVB-S2 code, and the remaining quantized bits are encoded with a second, similar, DVB-S2 code. The parity bits of these two encoders are then concatenated for transmission.



Our experiments involve 16-QAM at the normalized rate of 3.4 bits/s/Hz and 16-PSK at the rate of 3.27 bits/s/Hz. Each modulation is attached to a 4-level multi-level code, which in the case of 16-QAM have rates 0.9, 0.8, 0.9, and 0.8, and in the case of 16-PSK have rates 0.9, 0.9, 0.8, and 2/3.

\begin{figure}
\centering
\includegraphics[width=4.5in]{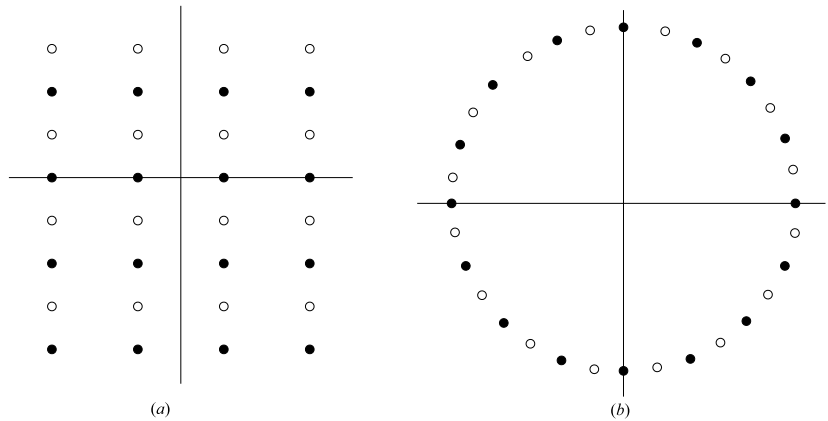}
\caption{The reconstruction values for CF relaying using 16-QAM (a), and 16-PSK (b), where the circles represent 16-QAM/PSK constellation.}
\label{fig:Reconstruction values}
\end{figure}

The relay utilizes a TCQ with an 8-state trellis whose generator matrix is:
\begin{equation*}
\begin{bmatrix}
1 & 0 & 0& 0& 0\\
0 & 1 & 0 & 0 & 0\\
0 & 0 & D & 1 & 0\\
0 & 0 & 1 & D^2 & D
\end{bmatrix}.
\end{equation*}
The quantizer reconstruction values, shown in Fig.~\ref{fig:Reconstruction values}, have a cardinality that is twice the size of the 16-QAM/PSK constellation used for source/relay modulation. For the 16-QAM example, extending to a standard 32AMPM would present inconvenient cell boundaries, therefore we experimented with the shown 32-point extension as well as a symmetric extension with 64 points, finding the latter produces negligible gains over the former. The design of quantizer bin boundaries is according to the optimization mentioned in Section~\ref{Sec:QuantizerOptimization}.


We compare our results with that of bit-interleaved coded modulation (BICM)~\cite{Nagpal_QMF_Allerton,Nagpal_QMF_Selected}, where the component LDPC codes in the sub-channels have the same rate by definition. Since the granularity of available DVB-2 codes are limited, we simulated BICM at the closest rate possible with available codes. A rate-5/6 DVB-S2 code is used for 16-QAM BICM, resulting in an overall rate of 3.33 bits/s/Hz, and a rate-0.8 DVB-S2 code was used for 16-PSK resulting in overall rate of 3.2 bits/s/Hz. In addition to the BICM results, we also display the results of a preliminary version of this work from~\cite{CF_Multilevel} which uses only scalar quantization. In addition, rates are carefully chosen to avoid any unfair advantage to the method of this letter in reporting error rates.

\begin{figure}
\centering
\includegraphics[width=\Figwidth]{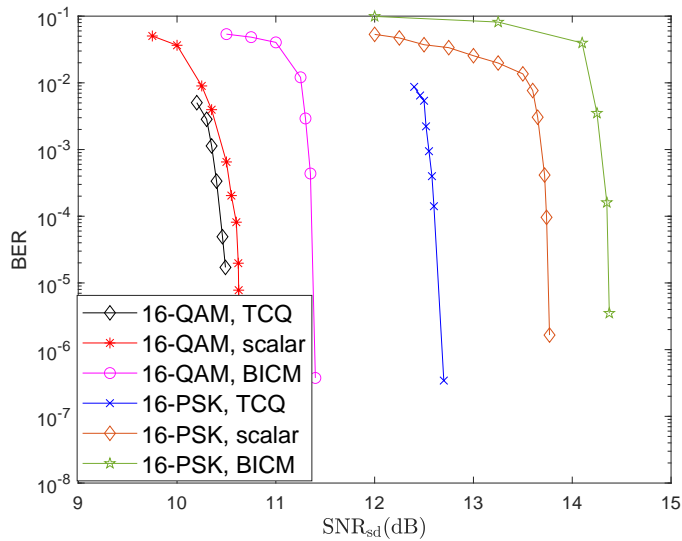}
\caption{Simulation results using 16-QAM/PSK with $N_3=1$, $N_2=8$, $H_{13}=1$, $H_{12}=2$, $H_{23}=11$, and $P_r=P_s$.}
\label{fig:BER}
\end{figure}

Simulation results are displayed in Fig.~\ref{fig:BER}. The observations and insights arising from these simulations are as follows:
\begin{itemize}
\item
The proposed method outperforms the error rate of BICM under 16-QAM and 16-PSK, even though the simulated BICM rates were slightly advantageous.
\item
Superior performance of MLC over BICM was known in point-to-point channels, but is affirmed in this letter for the end-to-end performance of a CF relaying.
\item
It is also established that TCQ can outperform scalar quantization in CF relaying. Although at first sight this may seem a natural outcome, past work~\cite{PracticalQuantizerDesign} implies that utilizing vector quantization in the context of CF has been remarkably difficult.  In that sense, the result reported herein has significant {\em practical} novelty.
\item
Approximately 1dB improvement is observed for 16-PSK. The improvement for 16-QAM is more modest.
\end{itemize}

\section{Discussion and conclusion}
\label{sec:con}
This paper significantly improves known implementations of compress-forward via a combination of trellis coded quantization at the relay, multi-level coding with LDPC component codes, and iterative joint decoding of the source and relay signals at the destination. The utilization of TCQ to facilitate the generation of bit LLRs at the destination is a key contribution of this work. Among others, this work shows a 1dB improvement for compress-forward with PSK modulations. 

\bibliographystyle{IEEEtran}
\balance
\bibliography{IEEEabrv,mybibfile}

\begin{thebibliography}{10}
\providecommand{\url}[1]{#1}
\csname url@samestyle\endcsname
\providecommand{\newblock}{\relax}
\providecommand{\bibinfo}[2]{#2}
\providecommand{\BIBentrySTDinterwordspacing}{\spaceskip=0pt\relax}
\providecommand{\BIBentryALTinterwordstretchfactor}{4}
\providecommand{\BIBentryALTinterwordspacing}{\spaceskip=\fontdimen2\font plus
\BIBentryALTinterwordstretchfactor\fontdimen3\font minus
  \fontdimen4\font\relax}
\providecommand{\BIBforeignlanguage}[2]{{%
\expandafter\ifx\csname l@#1\endcsname\relax
\typeout{** WARNING: IEEEtran.bst: No hyphenation pattern has been}%
\typeout{** loaded for the language `#1'. Using the pattern for}%
\typeout{** the default language instead.}%
\else
\language=\csname l@#1\endcsname
\fi
#2}}
\providecommand{\BIBdecl}{\relax}
\BIBdecl

\bibitem{Kramer_short}
G.~Kramer and J.~Hou, ``Short-message quantize-forward network coding,'' in
  \emph{International Workshop on Multi-Carrier Systems Solutions}, Herrsching,
  Germany, May 2011, pp. 1--3.

\bibitem{Zhong_withoutWynerZiv}
P.~Zhong and M.~Vu, ``Compress-forward without {Wyner-Ziv} binning for the
  one-way and two-way relay channels,'' in \emph{Proc. 49th Annual Allerton
  Conference on Communication, Control, and Computing}, Monticello, IL, USA,
  Sep. 2011, pp. 426--433.

\bibitem{Nagpal_QMF_Allerton}
V.~Nagpal, I.-H. Wang, M.~Jorgovanovic, D.~Tse, and B.~Nikoli\'c,
  ``Quantize-map-and-forward relaying: Coding and system design,'' in
  \emph{Allerton Conference on Communication, Control, and Computing},
  Allerton, IL, USA, Oct. 2010, pp. 443--450.

\bibitem{Nagpal_QMF_Selected}
V.~Nagpal, I.-H. Wang, and M.~Jorgovanovic, ``Coding and system design for
  quantize-map-and-forward relaying,'' \emph{IEEE J. Sel. Areas Commun.},
  vol.~31, no.~8, pp. 1423--1435, Aug. 2013.

\bibitem{A_layered_cf}
F.~U. Din, J.~N. Chattha, I.~Ullah, and M.~Uppal, ``A layered
  detect-compress-and-forward coding scheme for the relay channel,'' in
  \emph{IEEE International Symposium on Personal, Indoor, and Mobile Radio
  Communications (PIMRC)}, Oct. 2017, pp. 1--5.

\bibitem{Uppal_bpsk}
M.~Uppal, Z.~Liu, V.~Stankovic, and Z.~Xiong, ``Compress-forward coding with
  {BPSK} modulation for the half-duplex {Gaussian} relay channel,''
  \emph{{IEEE} Trans. Signal Process.}, vol.~57, no.~11, pp. 4467--4481, Nov.
  2009.

\bibitem{Uppal_MLC_Globecom}
J.~Amjad, M.~Uppal, and S.~Qaisar, ``Multi-level compress and forward coding
  for half-duplex relays,'' in \emph{Proc. IEEE Globecom}, Anaheim, CA, USA,
  Dec. 2012, pp. 4536--4541.

\bibitem{Blasco_polar_Ic}
R.~Blasco-Serrano, R.~Thobaben, M.~Andersson, V.~Rathi, and M.~Skoglund,
  ``Polar codes for cooperative relaying,'' \emph{{IEEE} Trans. Commun.},
  vol.~61, no.~2, pp. 3263--3273, Feb. 2015.

\bibitem{cf_superposition}
Y.~Liu, W.~B. Xu, K.~Niu, Z.~Q. He, and B.~Y. Tian, ``A practical
  compress-and-forward relay scheme based on superposition coding,'' in
  \emph{IEEE International Conference on Communication Technology}, Nov. 2012,
  pp. 1286--1290.

\bibitem{Infor_flow}
A.~S. Avestimehr, S.~N. Diggavi, and D.~N.~C. Tse, ``Wireless network
  information flow: A deterministic approach,'' \emph{{IEEE} Trans. Inf.
  Theory}, vol.~57, no.~4, pp. 1872--1905, Apr. 2011.

\bibitem{PracticalQuantizerDesign}
A.~Chakrabarti, A.~Sabharwal, and B.~Aazhang, ``Practical quantizer design for
  half-duplex estimate-and-forward relaying,'' \emph{{IEEE} Trans. Commun.},
  vol.~59, no.~1, pp. 74--83, Jan. 2011.

\bibitem{CF_Multilevel}
H.~{Wan}, A.~{Høst-Madsen}, and A.~{Nosratinia}, ``Compress-and-forward via
  multilevel coding,'' in \emph{{IEEE} Int. Symp. on Info. Theory (ISIT)}, Jul.
  2019, pp. 2024--2028.

\bibitem{TCQ}
M.~W. Marcellin and T.~R. Fischer, ``Trellis coded quantization of memoryless
  and gauss-markov sources,'' \emph{{IEEE} Trans. Commun.}, vol.~38, no.~1, pp.
  82--93, Jan. 1990.

\bibitem{BCJR_TrellisSourceCoding}
J.~B. Anderson, T.~Eriksson, and N.~Goertz, ``On the {BCJR} algorithm for
  rate-distortion source coding,'' \emph{{IEEE} Trans. Inf. Theory}, vol.~53,
  no.~9, pp. 3201--3207, Sep. 2007.

\bibitem{Imai_MLC_It}
H.~Imai and S.~Hirakawa, ``A new multilevel coding method using error
  correcting codes,,'' \emph{{IEEE} Trans. Inf. Theory}, vol.~23, no.~3, pp.
  371--377, May 1977.

\bibitem{ElGamalKim_NIT}
A.~El~Gamal and Y.-H. Kim, \emph{Network Information Theory}, 1st~ed.\hskip 1em
  plus 0.5em minus 0.4em\relax Cambridge, U.K: Cambridge University Press,
  2012.

\bibitem{TCM_Ungerboeck}
G.~Ungerboeck, ``Channel coding with multilevel/phase signals,'' \emph{{IEEE}
  Trans. Inf. Theory}, vol.~28, no.~1, pp. 55--67, Jan. 1982.

\bibitem{Shannon_RateDistortion}
C.~E. Shannon, ``Coding theorems for a discrete source with a fidelity
  criterion,'' \emph{IRE Nat. Conv. Rec.}, pp. 142--163, Mar. 1959.

\end{thebibliography}

\end{document}